# Quasiparticle poisoning rate in a superconducting transmon qubit involving Majorana zero modes

Xiaopei Sun [1,2,*], Zhaozheng Lyu[1,2,*,†], Enna Zhuo [1,2], Bing Li[1,2], Zhongqing Ji[1], Jie Fan[1], Xiaohui Song[1,4],
Fanning Qu[1,2,3,4], Guangtong Liu[1,2,3,4], Jie Shen[1,2,3], and Li Lu[1,2,3,4,†]

[1]*Beijing National Laboratory for Condensed Matter Physics, Institute of Physics, Chinese Academy of Sciences, Beijing 100190, China*

[2]*School of Physical Sciences, University of Chinese Academy of Sciences, Beijing 100049, China*

[3]*Songshan Lake Materials Laboratory, Dongguan, Guangdong 523808, China*

[4]*Hefei National Laboratory, Hefei 230088, China*

Majorana zero modes have been attracting considerable attention because of their prospective applications in fault-tolerant topological quantum computing. In recent years, some schemes have been proposed to detect and manipulate Majorana zero modes using superconducting qubits. However, manipulating and reading the Majorana zero modes must be kept in the time window of quasiparticle poisoning. In this work, we study the problem of quasiparticle poisoning in a split transmon qubit containing hybrid Josephson junctions involving Majorana zero modes. We show that Majorana coupling will cause parity mixing and $4\pi$ Josephson effect. In addition, we obtained the expression of qubit parameter-dependent parity switching rate, and demonstrated that quasiparticle poisoning can be greatly suppressed by reducing $E_J/E_C$ via qubit design.

## I. INTRODUCTION

There are many theoretical and experimental explorations along the research direction of realizing fault-tolerant topological quantum computation [1–6]. Among them, a highly regarded scheme is to combine circuit quantum electrodynamics (cQED) with Josephson junctions (JJs) involving Majorana zero modes (MZMs) [7–13]. It is theoretically proposed that a superconducting qubit, such as the Majorana transmon qubit in Ref. [12], can be used to detect and manipulate MZMs. Experimentally, hybrid superconducting qubits have been realized by replacing the superconductor-insulator-superconductor (S-I-S) JJs in conventional superconducting qubits with the JJs based on a variety of materials, such as semiconductor [14–16], van der Waals materials [17,18], and topological materials [19,20]. These hybrid qubits provide a platform for studying Majorana physics and topological quantum computation, since MZMs could be found in the heterojunctions between these materials and s-wave superconductors.

However, both Majorana qubits and superconducting qubits are susceptible to quasiparticle poisoning (QPP) [21–29]. Although the problem of QPP in conventional S-I-S superconducting qubits has been well studied both theoretically [24–27] and experimentally [28–33], QPP in superconducting qubits involving MZMs remains unexplored.

In this work, we consider a split transmon qubit containing MZMs, as schematically depicted in Fig. 1. The parity mixing effect and the $4\pi$ Josephson effect caused by Majorana coupling are found in the calculated excitation spectrum of the qubit. We further demonstrate that QPP can be greatly suppressed by reducing $E_J/E_C$ via qubit design.

*These authors contribute equally to this work.

† Corresponding authors. Zhaozheng Lyu: lyuzhzh@iphy.ac.cn,
Li Lu: lilu@iphy.ac.cn

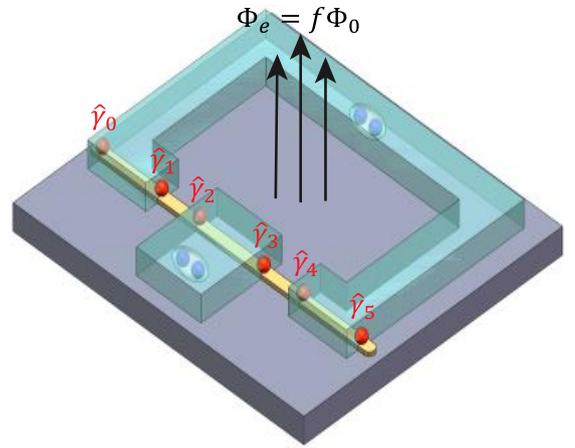

Fig. 1. Schematic of a split transmon qubit with MZMs. Two superconductors (cyan-blue) on a topological nanowire (yellow) form a split transmon with six MZMs (red) in the device.

## II. THE TOTAL HAMILTONIAN OF THE MAJORANA-TRANSMON QUBIT

The total hybrid Hamiltonian describing the device in Fig. 1 can be split into four parts [12,25]:

$$\hat{H} = \hat{H}_\phi + \hat{H}_M + \hat{H}_{qp} + \hat{H}_T. \tag{1}$$

The first term describes the regular split transmon Hamiltonian

$$\hat{H}_\phi = 4E_C(\hat{N} - n_g)^2 - E_J(f)\cos\hat{\phi}, \qquad f = \frac{\Phi_e}{\Phi_0}, \tag{2}$$

where $\hat{N}$ is the number operator of Cooper pairs, $\hat{\phi}$ is the phase operator, $n_g$ is the dimensionless gate voltage, $\Phi_0 = h/2e$ is the flux quantum, $E_C$ is the charging energy, and the effective Josephson energy $E_J(f)$ is modulated by the external flux $\Phi_e$:

$$E_J(f) = (E_{J0} + E_{J1})|\cos(\pi f)|\sqrt{1 + d^2\tan^2(\pi f)}, \tag{3}$$

where $E_{J0}$ and $E_{J1}$ are the Josephson energy of two junctions respectively, and $d = (E_{J0} - E_{J1})/(E_{J0} + E_{J1})$ is the junction asymmetry parameter. The qubit frequency of $\hat{H}_\phi$ is approximately $\omega_p(f) = \sqrt{8E_C E_J(f)}$.

The second term determines the couplings between the MZMs distributed across the Josephson junctions

$$\hat{H}_M = 2iE_{M0}\hat{\gamma}_1\hat{\gamma}_2 \cos\left(\frac{\hat{\varphi}_0}{2}\right) + 2iE_{M1}\hat{\gamma}_3\hat{\gamma}_4 \cos\left(\frac{\hat{\varphi}_1}{2}\right), \quad (4)$$

where $\hat{\gamma}_1, \hat{\gamma}_2, \hat{\gamma}_3$ and $\hat{\gamma}_4$ are MZMs operators as shown in Fig. 1, $E_{M0}$ and $E_{M1}$ are coupling strengths between the MZMs distributed across the Josephson junctions, $\hat{\varphi}_0$ and $\hat{\varphi}_1$ describe the phase difference of two junctions respectively:

$$\begin{cases} \hat{\varphi}_0 = \pi f - \vartheta - \hat{\phi} \\ \hat{\varphi}_1 = \pi f + \vartheta + \hat{\phi} \end{cases}, \quad \tan\vartheta = d \tan(\pi f) \quad (5)$$

The couplings between MZMs on the same electrodes have been ignored in Eq. (4) since they only cause small corrections in the diagonal elements and do not affect our results.

The third term is the sum of the BCS Hamiltonians for quasiparticles in the leads

$$\hat{H}_{qp} = \sum_{j=0,1} \hat{H}_{qp}^j, \quad \hat{H}_{qp}^j = \sum_{n,\sigma} \epsilon_n^j \hat{\alpha}_{n\sigma}^{j\dagger} \hat{\alpha}_{n\sigma}^j, \quad (6)$$

where the index $j$ denotes the superconducting island, $\hat{\alpha}_{n\sigma}^{j\dagger}$ ($\hat{\alpha}_{n\sigma}^j$) are Bogoliubov quasiparticle creation (annihilation) operators with the spin $\sigma$, and $\epsilon_n^j = \sqrt{(\xi_n^j)^2 + (\Delta^j)^2}$ are quasiparticle energies with $\xi_n^j$ and $\Delta^j$ being the single-particle energy level and the superconducting gap in that lead, respectively. In the following we assume equal gaps in the leads $\Delta^j \simeq \Delta$ for simplicity.

The last item is the tunnel Hamiltonian $\hat{H}_T$ which describes quasiparticle tunneling across the junction

$$\begin{aligned} \hat{H}_T &= \sum_{j=0,1} \tilde{t}_j \sum_{n,m,\sigma} \left( e^{i\frac{\hat{\varphi}_j}{2}} u_n^j u_m^{j+1} - e^{-i\frac{\hat{\varphi}_j}{2}} v_m^{j+1} v_n^j \right) \hat{\alpha}_{n\sigma}^{j\dagger} \hat{\alpha}_{m\sigma}^{j+1} + \text{H.c.} \\ &\simeq \sum_{j=0,1} \tilde{t}_j \sum_{n,m,\sigma} i \sin\frac{\hat{\varphi}_j}{2} \hat{\alpha}_{n\sigma}^{j\dagger} \hat{\alpha}_{m\sigma}^{j+1} + \text{H.c.} \end{aligned} \quad (7)$$

where the island $j = 2$ should be identified as the island $j = 0$, $\tilde{t}_j$ are electron tunneling amplitudes between islands $j$ and $j + 1$. $\hat{H}_T$ leads to the interaction between quasiparticles and qubit degrees of freedom and makes possible the transition between two qubit states induced by QPP. Since the tunneling amplitude is usually small $\tilde{t}_j \ll 1$, we can calculate the transition rates by treating $\hat{H}_T$ as a perturbation term.

## III. THE SOLUTION OF THE HYBRID QUBIT SYSTEM

In contrast to the regular transmon qubits, the Majorana coupling term $\hat{H}_M$ in Eq. (4) will double the degeneracy of the qubit states, and the charge parity of the qubit will be jointly determined by the quasiparticle parity and the Majorana parity. The Hamiltonian in Eq. (1) can be projected on four different parity states:

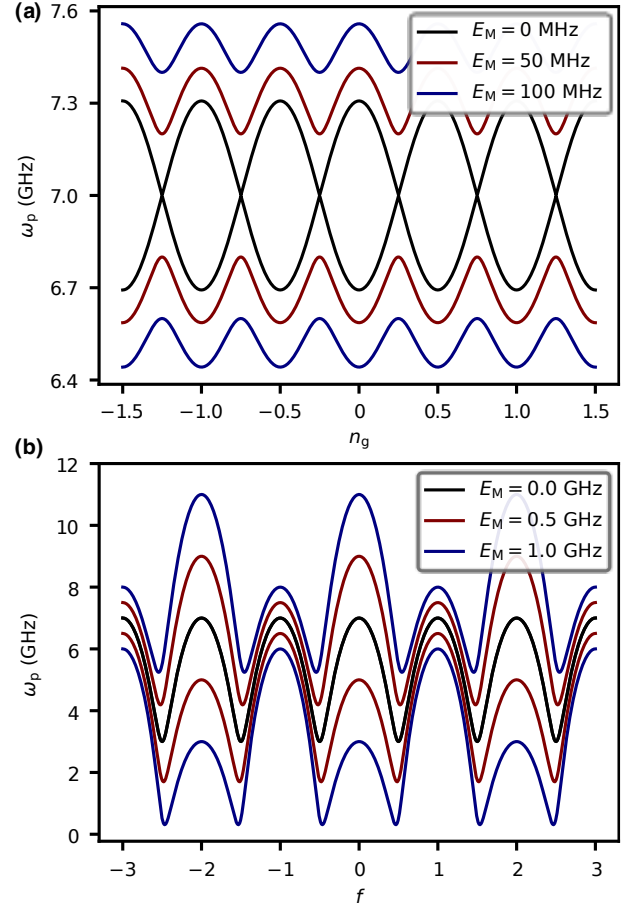

Fig. 2. Calculated excitation spectrum of the Majorana transmon qubit with $E_C = 1\text{ GHz}, E_J = 8\text{ GHz}$, and $d = 0.25$. For simplicity, we denote the average of $E_{M1}$ and $E_{M2}$ as $E_M = (E_{M1} + E_{M2})/2$. (a) The excitation spectrum as a function of gate voltage $n_g$ with $E_M = 0$ (black), $E_M = 50$ MHz (red), and $E_M = 100$ MHz (blue). (b) The excitation spectrum as a function of flux $f$ with $E_M = 0$ (black), $E_M = 0.5$ GHz (red), and $E_M = 1$ GHz (blue).

$$\begin{pmatrix} |\psi_n^e, \{e\}_{qp}, \{e\}_M\rangle \\ |\psi_n^o, \{e\}_{qp}, \{o\}_M\rangle \\ |\psi_n^e, \{o\}_{qp}, \{o\}_M\rangle \\ |\psi_n^o, \{o\}_{qp}, \{e\}_M\rangle \end{pmatrix} \quad (8)$$

where $\psi_n^\zeta$ represents the eigenstate of transmon, $\zeta$ denotes the total charge parity, $n$ denotes the energy level of the qubit, $\{e\}_{qp}$ ($\{o\}_{qp}$) corresponds to the even (odd) states of the quasiparticles, $\{e\}_M$ consists of the Majorana parity states $|01\rangle$ and $|00\rangle$, $\{o\}_M$ consists of the Majorana parity states $|10\rangle$ and $|11\rangle$. By replacing the Majorana operators in Eq. (4) with the fermion operators (e.g., $-2i\hat{\gamma}_1\hat{\gamma}_2 = \hat{c}_0^\dagger \hat{c}_1 + \hat{c}_1^\dagger \hat{c}_0 + \hat{c}_0^\dagger \hat{c}_1^\dagger + \hat{c}_1 \hat{c}_0$), it is straightforward to find that the matrix elements of $\hat{H}_M$ are non-zero only when the two states have different Majorana parities but the same quasiparticle parity

$$\langle \psi_n^\zeta, \{\lambda\}_{qp}, \{\eta\}_M | \hat{H}_M | \psi_n^{\bar{\zeta}}, \{\lambda\}_{qp}, \{\bar{\eta}\}_M \rangle = \sum_{j=0,1} E_{Mj} \langle \psi_n^\zeta | \cos\frac{\hat{\varphi}_j}{2} | \psi_n^{\bar{\zeta}} \rangle. \quad (9)$$

In the transmon limit $E_J/E_C \gg 1$, we can expand Eq. (2) around $\phi = 0$ up to the second order, and then use the standard harmonic oscillator eigenstates as the qubit states. In order to obtain the analytic expressions for the matrix elements of $\cos(\hat{\varphi}_j/2)$, we introduce the displacement operator

$$\hat{D}(\mu) = e^{\mu \hat{a}^\dagger - \mu^* \hat{a}}, \qquad (10)$$

where $\hat{a}$ and $\hat{a}^\dagger$ are the annihilation and creation operators respectively. When $\mu = i\sqrt{E_C/\omega_p}$, by using $\hat{\phi} = 2\sqrt{E_C/\omega_p}(\hat{a} + \hat{a}^\dagger)$, we have

$$\hat{D}\left(i\sqrt{E_C/\omega_p}\right) = e^{i\hat{\phi}/2}. \qquad (11)$$

Since $\hat{\varphi}_j$ is shifted by $\varphi_\pm = \vartheta \pm \pi f$ in Eq. (S6), $\cos(\hat{\varphi}_j/2)$ can be expressed as

$$\begin{aligned}\cos\frac{\varphi_\pm + \hat{\phi}}{2} &= \frac{1}{2}\left(e^{i\varphi_\pm/2}e^{i\hat{\phi}/2} + e^{-i\varphi_\pm/2}e^{-i\hat{\phi}/2}\right) \\ &= \frac{1}{2}\left[e^{\frac{i\varphi_\pm}{2}}\hat{D}\left(i\sqrt{\frac{E_C}{\omega_p}}\right) + e^{-\frac{i\varphi_\pm}{2}}\hat{D}\left(-i\sqrt{\frac{E_C}{\omega_p}}\right)\right].\end{aligned} \qquad (12)$$

The matrix elements of $\hat{D}(\mu)$ are [34]:

$$\langle \psi_m | \hat{D}(\mu) | \psi_n \rangle = \begin{cases} e^{-|\mu|^2/2}\sqrt{\dfrac{m!}{n!}}(-\mu^*)^{n-m}L_m^{(n-m)}(|\mu|^2), & m \leq n \\[6pt] e^{-|\mu|^2/2}\sqrt{\dfrac{n!}{m!}}(\mu)^{m-n}L_n^{(m-n)}(|\mu|^2), & m \geq n \end{cases} \qquad (13)$$

where $L_n^{(l)}$ are the generalized Laguerre polynomials and can be expanded as

$$L_m^{(l)}(x) = \frac{(m+l)!}{m!l!} - \frac{(m+l)!}{(m-1)!(l+1)!}x + O(x^2). \qquad (14)$$

Since $\mu = i\sqrt{E_C/\omega_p} = i(E_C/8E_J)^{1/4} \ll 1$ in the transmon limit $E_J/E_C \gg 1$, we can expand Eq. (12) up to the second order of $\mu$

$$\begin{aligned}\langle \psi_m | \cos\frac{\hat{\varphi}_j}{2} | \psi_n \rangle &= \\ &\left[1 - \left(n + \frac{1}{2}\right)\frac{E_C}{\omega_p}\right]\delta_{m,n}\cos\frac{\hat{\varphi}_\pm}{2} \\ &- \sqrt{\frac{E_C}{\omega_p}}\left[n\delta_{m,n-1} + (n+1)\delta_{m,n+1}\right]\sin\frac{\hat{\varphi}_\pm}{2} \\ &+ \frac{1}{2}\frac{E_C}{\omega_p}\left[\sqrt{n(n-1)}\delta_{m,n-2} + \sqrt{(n+1)(n+2)}\delta_{m,n+2}\right]\cos\frac{\hat{\varphi}_\pm}{2}.\end{aligned} \qquad (15)$$

For the qubit degrees of freedom we are considering, the quasiparticle term $\hat{H}_{qp}$ can be neglected. And we consider first the solution at ground states and then generalize the result to higher energies. In the basis of Eq. (8), the hybrid Hamiltonian yields the form

$$\hat{H}_Q = \hat{H}_\phi + \hat{H}_M = \frac{1}{2}\begin{pmatrix} \omega_{eo} & \omega_M & & \\ \omega_M & -\omega_{eo} & & \\ & & \omega_{eo} & \omega_M \\ & & \omega_M & -\omega_{eo} \end{pmatrix}, \qquad (16)$$

where $\omega_M = 2\langle \psi_0^\zeta, \{\lambda\}_{qp}, \{\eta\}_M | \hat{H}_M | \psi_0^{\bar{\zeta}}, \{\lambda\}_{qp}, \{\bar{\eta}\}_M \rangle$, substituting Eq. (15) into Eq. (9), we obtain

$$\omega_M \simeq 2\sum_{j=0,1} E_{Mj} \cos\frac{\pi f - (-1)^j \vartheta}{2}, \qquad (17)$$

$\omega_{eo}$ is the energy difference between even and odd ground states of the regular transmon. Following Appendix B of Ref. [25], $\omega_{eo}$ can be expressed as

$$\omega_{eo} = \epsilon_0 \cos(2\pi n_g) \simeq 4\sqrt{\frac{2}{\pi}}\omega_p \left(\frac{8E_J}{E_C}\right)^{\frac{1}{4}} e^{-\sqrt{8E_J/E_C}} \cos(2\pi n_g). \qquad (18)$$

The two states corresponding to the eigenvalue $E_0^+ = \omega_{eo}'/2 = \sqrt{\omega_{eo}^2 + \omega_M^2}/2$ are:

$$\begin{cases} |\Psi_0^+, \{e\}_{qp}\rangle = a^+|\psi_0^e, \{e\}_{qp}, \{e\}_M\rangle + b^+|\psi_0^o, \{e\}_{qp}, \{o\}_M\rangle \\ |\Psi_0^+, \{o\}_{qp}\rangle = a^+|\psi_0^e, \{o\}_{qp}, \{o\}_M\rangle + b^+|\psi_0^o, \{o\}_{qp}, \{e\}_M\rangle \end{cases} \qquad (19)$$

and the two states corresponding to the eigenvalue $E_0^- = -\omega_{eo}'/2 = -\sqrt{\omega_{eo}^2 + \omega_M^2}/2$ are:

$$\begin{cases} |\Psi_0^-, \{e\}_{qp}\rangle = a^-|\psi_0^e, \{e\}_{qp}, \{e\}_M\rangle + b^-|\psi_0^o, \{e\}_{qp}, \{o\}_M\rangle \\ |\Psi_0^-, \{o\}_{qp}\rangle = a^-|\psi_0^e, \{o\}_{qp}, \{o\}_M\rangle + b^-|\psi_0^o, \{o\}_{qp}, \{e\}_M\rangle \end{cases} \qquad (20)$$

where

$$\begin{aligned} a^\pm &= \frac{\omega_{eo} \pm \omega_{eo}'}{\sqrt{\omega_M^2 + (\omega_{eo} \pm \omega_{eo}')^2}}, \\ b^\pm &= \frac{\omega_M}{\sqrt{\omega_M^2 + (\omega_{eo} \pm \omega_{eo}')^2}}. \end{aligned} \qquad (21)$$

$\omega_{eo}' = \sqrt{\omega_{eo}^2 + \omega_M^2}$ is the energy splitting between the hybrid ground states, which indicates an energy splitting of $\omega_M$ at the parity crossing at $n_g = 1/4$. This result agrees well with the calculated spectrum in Ref. [12], thus validating our approach. Following Ref. [25], the solution at the excited state has a similar expression to Eq. (19-21) but with $\omega_{eo}$ replaced by the energy splitting at the excited state $\omega_{eo}^{|1)} = -4\omega_{eo}\omega_p/E_C$. Figure 2(a) shows the energy splitting in excitation spectrum caused by the parity mixing effect and Fig. 2(b) shows the $4\pi$ Josephson effect. Both these characteristics are crucial criteria for identifying MZMs.

## IV. THE PARITY-SWITCHING RATES

Next, we calculate the parity switching rate at the ground state of the hybrid system by using Fermi's golden rule and treating $\hat{H}_T$ as a perturbation term, following Ref. [25], we can write

$$\begin{aligned}\Gamma_{00}^{oe} = 2\pi \sum_{\kappa'=\pm} \sum_{\{e\}_{qp}} \Big\langle \Big| \Big|\langle \Psi_0^{\kappa'}, \{e\}_{qp} | \hat{H}_T | \Psi_0^\kappa, \{o\}_{qp}\rangle \Big|^2 \\ \times \delta(E_{e,qp} - E_{o,qp} - \omega_{eo}') \Big\rangle \Big\rangle_{\kappa, \{o\}_{qp}}\end{aligned} \qquad (22)$$

where $E_{o,qp}$ ($E_{e,qp}$) is the total energy of quasiparticles in their initial (final) state $\{o\}_{qp}$ ($\{e\}_{qp}$), and the double angular brackets $\langle\langle \cdots \rangle\rangle_{\kappa,\{o\}_{qp}}$ denote averaging over the initial quasiparticle states. In the low-energy regime, substituting Eq. (7), Eq. (19) and Eq. (20) into Eq. (22) and using $\tilde{t}_j^2 \propto E_{Jj}$, we can factorize $\Gamma_{00}^{oe}$ into terms accounting separately for qubit dynamic and quasiparticle kinetics

$$\begin{aligned}\Gamma_{00}^{oe} &= \sum_{j=0,1} \frac{E_{Jj}}{E_J(f)} \left|\langle\psi_0^e|\sin\frac{\hat{\varphi}_j}{2}|\psi_0^o\rangle\right|^2 \\
&\quad \times [(2a^+b^+)^2 S_{qp}(0) + (a^+b^- + a^-b^+)^2 S_{qp}(\omega'_{eo})] \\
&= \sum_{j=0,1} \frac{E_{Jj}}{E_J(f)} \left|\langle\psi_0^e|\sin\frac{\hat{\varphi}_j}{2}|\psi_0^o\rangle\right|^2 \\
&\quad \times \left[\frac{\omega_M^2}{\omega_M^2 + \omega_{eo}^2} S_{qp}(0) + \frac{\omega_{eo}^2}{\omega_M^2 + \omega_{eo}^2} S_{qp}(\omega'_{eo})\right] \\
&= \frac{E_{J0} + E_{J1}}{2E_J(f)}\left(1 - \frac{\omega_p^2(f)}{\omega_p^2(0)}\right) \\
&\quad \times \left[\frac{\omega_M^2}{\omega_M^2 + \omega_{eo}^2} S_{qp}(0) + \frac{\omega_{eo}^2}{\omega_M^2 + \omega_{eo}^2} S_{qp}(\omega'_{eo})\right]\end{aligned} \quad (23)$$

where, going from the second to the third line, we calculate the matrix elements of $\sin(\hat{\varphi}_j/2)$ in a similar way to the calculation of Eq. (15), and $S_{qp}$ is the quasiparticle current spectral density which can be expressed as{Citation}:

$$S_{qp}(\omega) = \frac{16 E_J(f)}{\pi} e^{-\Delta/T} e^{\omega/2T} K_0\left(\frac{|\omega|}{2T}\right), \quad (24)$$

where $T$ is the temperature, $K_0$ is the modified Bessel function of the second kind.

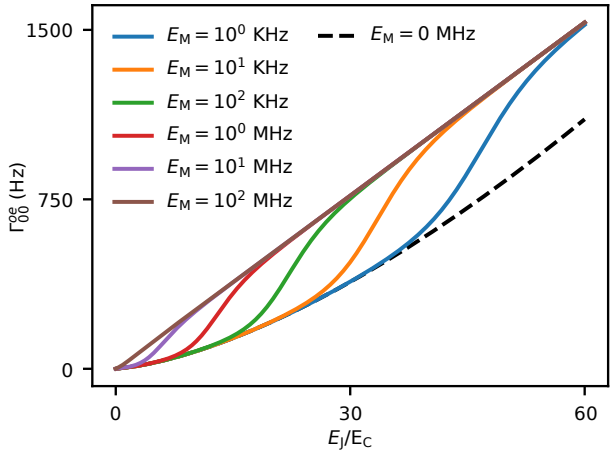

Fig. 3. The parity-switching rate at the ground state of the symmetric split transmon as a function of $E_J/E_C$ with $E_C = 0.2$ GHz, $n_g = 0$, $f = 0.25$, $T = 100$ mK, $\Delta = 161$ μeV and for different $E_M$. The black dashed line corresponds to the regular split transmon (i.e. $E_M = 0$).

We remind that $\omega'_{eo} = \sqrt{\omega_{eo}^2 + \omega_M^2}$. It is straightforward to find that when $\omega_M = 0$, Eq. (23) is in agreement with the result of the regular split transmon in Ref. [25]. The parity-switching rate at the excited state $\Gamma_{11}^{oe}$ can also be calculated by a similar way, and has a similar expression to Eq. (23) in the transmon limit but with $\omega_{eo}$ replaced by the energy splitting at the excited state $\omega_{eo}^{|1\rangle} = -4\omega_{eo}\omega_p/E_C$.

Since $\omega_{eo}$ is exponentially suppressed by the increasing $E_J/E_C$ as shown in Eq. (18), the parity-switching rates at the same qubit states (i.e., $\Gamma_{00}^{oe}$ and $\Gamma_{11}^{oe}$) are strongly correlated with the qubit design $E_J/E_C$, as shown in Fig. 3. Therefore, we can suppress $\Gamma_{00}^{oe}$ and $\Gamma_{11}^{oe}$ by reducing $E_J/E_C$ via qubit design.

## V. SUMMARY

In this work, we have studied the QPP in a split transmon qubit involving Majorana-based Josephson effect. By approximating the transmon qubit as a harmonic oscillator, we have solved the Hamiltonian of this hybrid system analytically and demonstrated the parity mixing effect and the $4\pi$ Josephson effect caused by Majorana coupling. These two phenomena could provide an important evidence for MZMs in microwave experiments. In addition, we have derived the expression of qubit parameter-dependent parity-switching rate, and showed that QPP can be greatly suppressed by reducing $E_J/E_C$ via qubit design.

## ACKNOWLEDGMENTS

This work was supported by the National Basic Research Program of China through MOST Grant Nos. 2016YFA0300601, 2017YFA0304700 and 2015CB921402; by NSFC through Grant Nos. 11527806, 92065203, 12074417, 11874406 and 11774405; by the Strategic Priority Research Program B of the Chinese Academy of Sciences through Grants Nos. XDB33010300, DB28000000 and XDB07010100; by the Synergetic Extreme Condition User Facility (SECUF) which is sponsored by the National Development and Reform Commission; by the NSFC for Young Scholars through Grant No. E2J1141; and by the Innovation Program for Quantum Science and Technology through Grant No. 2021ZD302600.